\documentclass[11pt]{article}  

\usepackage{amssymb,amstext,amsmath,amsthm}
\usepackage[dvips]{graphicx}
\usepackage{latexsym}
\usepackage{psfrag}
\usepackage{amsfonts}
\usepackage{bbm}
\usepackage{color}

\newcommand{\Ref}[1]{(\ref{#1})}

\newcommand{\be}{\begin{equation}}
\newcommand{\ee}{\end{equation}}
\newcommand{\barray}{\begin{array}}
\newcommand{\earray}{\end{array}}
\newcommand{\bea}{\begin{eqnarray}}
\newcommand{\eea}{\end{eqnarray}}
\newcommand{\bs}{\begin{subequations}}
\newcommand{\es}{\end{subequations}}
\newcommand{\balign}{\begin{align}}
\newcommand{\ealign}{\end{align}}
\newcommand{\equ}{\begin{equation}}
\newcommand{\nequ}{\end{equation}}
\newcommand{\eqa}{\begin{eqnarray}}
\newcommand{\neqa}{\end{eqnarray}}
\def\nn{\nonumber}
\newcommand{\vj}{\vec{\jmath} }

\def\w{\wedge}

\newcommand{\p}{\partial}
\newcommand{\n}{\nabla}

\makeatletter
\newcommand*{\simboloG}[1]{%
  \vphantom{\sum}
  \smash{%
    \mathchoice{%
      \raisebox{-.3\height}{\Huge$\m@th\displaystyle#1$}%
      }
      {%
      \raisebox{-.1\height}{\Large$\m@th#1$}%
      }{%
      \raisebox{-.1\height}{\small$\m@th#1$}%
      }{%
      \raisebox{-.1\height}{\LARGE$\m@th#1$}%
      }%
    }}
\newcommand{\BigTimes}{\mathop{\simboloG{\times}}}
\makeatother
\makeatletter

\newcommand*{\simboloB}[1]{%
  \vphantom{\sum}
  \smash{%
    \mathchoice{%
      \raisebox{-.1\height}{\Large$\m@th\displaystyle#1$}%
      }
      {%
      \raisebox{-.1\height}{\Large$\m@th#1$}%
      }{%
      \raisebox{-.1\height}{\small$\m@th#1$}%
      }{%
      \raisebox{-.1\height}{\LARGE$\m@th#1$}%
      }%
    }}
\newcommand{\bigTimes}{\mathop{\simboloB{\times}}}
\makeatother

\newcommand{\Z}{\mathbb{Z}}

\newcommand{\R}{\mathbb{R}}
\DeclareMathOperator{\tr}{Tr}

\newcommand{\cs}{{ S}}

\def\to{\rightarrow}

\def\d{\delta}
\def\f{\frac}
\def\tl{\tilde}

\def\Ga{\Gamma}
\def\ga{\gamma}

\usepackage{bbm}

\def\R{{\mathbbm R}}
\def\Z{{\mathbbm Z}}
\newcommand{\scr}{\rm\scriptscriptstyle}

\newcommand{\SU}{\mathrm{SU}}

\newcommand{\LX}{{\cal L}_{\hat X}}

\let\eps=\epsilon
\def\th{\theta}
\def\al{\alpha}

\newcommand{\tN}{\tl{N}}

\newcommand{\bz}{\bar{z}}

\newcommand{\rd}{\mathrm{d}}
\newcommand{\OmT}{\Omega_{\resizebox{!}{6pt}{$T^*{S^1}$}}}
\newcommand{\OmS}{\Omega_{\resizebox{!}{6pt}{$S^2$}}}
\newcommand{\ThT}{\Theta_{\resizebox{!}{6pt}{$T^*{S^1}$}}}
\newcommand{\ThS}{\Theta_{\resizebox{!}{6pt}{$S^2$}}}

\newcommand{\bA}{{\bf A}}
\newcommand{\bE}{{\bf E}}
\newcommand{\bG}{{\bf \Ga}}
\newcommand{\bK}{{\bf K}}

\begin{document}

\title{ Twisted geometries: \\A geometric parametrisation of SU(2) phase space
}
\author{{Laurent Freidel${}^{a}$ and Simone Speziale${}^{b}$}\footnote{lfreidel@perimeterinstitute.ca, simone.speziale@cpt.univ-mrs.fr}
\\
{\small ${}^a$\emph{Perimeter Institute for Theoretical Physics, 31 Caroline St. N, ON N2L 2Y5, Waterloo,Canada}} \\ 
{\small ${}^b$\emph{Centre de Physique Th\'eorique,\footnote{Unit\'e Mixte de Recherche (UMR 6207) du CNRS et des Universites Aix-Marseille I, Aix-Marseille II et du Sud Toulon-Var. Laboratoire affili\'e \`a la FRUMAM (FR 2291).} CNRS-Luminy Case 907, 13288 Marseille Cedex 09, France}}}
\date{\today}

\maketitle

\begin{abstract}
A cornerstone of the loop quantum gravity program is the fact that the phase space of general relativity on
a fixed graph can be described by a product of SU(2) cotangent bundles per edge. 
In this paper we show how to parametrize this phase space in terms of quantities describing the intrinsic and extrinsic geometry of the triangulation dual to the graph. 
These are defined by the assignment to each face of its area, the two unit normals as seen from the two polyhedra sharing it, and an additional angle related to the extrinsic curvature. These quantities do not define a Regge geometry, since they include extrinsic data, but a looser notion of discrete geometry which is \emph{twisted} in the sense that it is locally well-defined, but the local patches lack a consistent gluing among each other.
We give the Poisson brackets among the new variables, and exhibit a symplectomorphism which maps them into the Poisson brackets of loop gravity. The new parametrization has the  advantage of a simple description of the gauge-invariant reduced phase space, which is given by a product of phase spaces associated to edges and vertices, and it also provides an abelianisation of the SU(2) connection.
The results are relevant for the construction of coherent states, and as a byproduct, contribute to clarify the connection between loop gravity and its subset corresponding to Regge geometries.
\end{abstract}

\section{Introduction}

Spin network states are the building blocks of loop quantum gravity. They provide a basis of the kinematical Hilbert space  which diagonalizes some geometric operators, such as surface areas. The resulting discrete spectra, with minimal excitation proportional to the Planck length, are a landmark of the whole approach. In spite of the key role played in the theory, spin network states lack a low-energy physical interpretation. How can we bridge from the Planck scale quantum geometry they describe, to a smooth and classical three dimensional geometry?
To answer this question, one is interested in the construction of coherent states, namely superpositions of spin networks peaked on classical geometries labeling the phase space of the theory. The phase space of general relativity is parametrized by a canonically conjugated pair $(g_{ab}, K_{ab})$, the metric and extrinsic curvature of a three dimensional spacelike manifold, or equivalently by $(E_i^a, A_a^i)$, the triad and SU(2) connection used in loop gravity.
Therefore a coherent state for loop gravity should be a superposition of spin networks peaked on $(E_i^a, A_a^i)$, with small fluctuations.

After an early attempt with the weave states \cite{weave}, an extensive approach to this question has been developed by Thiemann and collaborators \cite{Thiemann, Bahr, Flori}. While the weave states only approximate either the connection or the triad field, Thiemann's coherent states are peaked on both with small fluctuations. The states are superpositions of spin networks with the same graph, and are properly labeled by a point in the discrete phase space of loop gravity associated to the graph.
Thiemann's states fulfill a number of important properties, including reproducing some aspects of the classical Hamiltonian constraint. 
On the other hand, the spin foam graviton calculations \cite{grav,gravEPR} and recent progress in the spin foam formalism \cite{EPR,LS,LS2,FK,CF1,CF2,CF3,Barrett}, suggest a different approach to the question, 
where one attempts at describing a classical phase space point in terms of quantities referring to discrete geometries, e.g. areas and dihedral angles, as opposed to holonomies and fluxes. 
To make this idea work, one needs  $(i)$ to show that the \emph{complete} phase space of the theory can be described by the candidate labels, and $(ii)$ to construct explicitly the associated coherent states and show that they satisfy the right properties, such as peakedness and (over)completeness.

In this paper we focus on step $(i)$. 
So although our final goal has to do with the quantum theory, here we deal entirely with the classical theory. 
The results of the paper  will be fundamental to take step $(ii)$. In a forthcoming paper \cite{noiQ}, we will construct explicitly the coherent states and compare them with Thiemann's and with the ones conjectured in the spin foam graviton calculations, in particular the ones recently appeared in \cite{BianchiCS} which bear similarities with our approach.

\subsection{Phase space of loop gravity}

The continuum phase space of loop gravity is defined by the Ashtekar-Barbero connection $A_a^i$ and the triad field $E^a_i$, satisfying the Poisson algebra
\be\label{AE}
\left\{ A^i_a(x), E^b_j(y)\right\} = \ga \, \d^i_j \d^b_a \d^{(3)}(x,y),
\ee
where $\ga$ is the Immirzi parameter. 
The connection with the ADM phase space of general relativity is established thanks to the fact that the Ashtekar-Barbero connection splits as
\equ\label{Aconn}
A_a = \Ga_a(E) + {\ga} \, K_a \in \mathfrak{su(2)},
\nequ
where $\Ga(E)$ is the canonical slice of the spin connection and $K$ the extrinsic curvature. 
Thanks to this key splitting, \Ref{AE} reduces to the ADM Poisson brackets, and the classical theory to general relativity for the metric associated to $E$. 

An important step towards quantization is the smearing of the algebra \Ref{AE}. This is done through the introduction of a 
graph $\Ga$ embedded in the spatial manifold, and replacing $(A^i_a, E^a_i)$ by a pair $(g_e,X_e)\in \SU(2)\times\mathfrak{su(2)}$ on each edge. These variables represent the connection and triad respectively as holonomies, $g_e = {\cal P} \exp \int_e A$ where $\cal P$ denotes the path-ordered product, and fluxes $X_e = \int_{e^*} (g E)^a N_a \rd^2S$ where $e^*$ is the dual face to the edge, with normal $N_a$ and infinitesimal coordinate area $\rd^2S$ and $g$ is the parallel transport from one fixed vertex to the point of integration on a path adapted to the graph. Since $\SU(2)\times{\mathfrak{su(2)}}\cong T^*\SU(2)$, we see that the phase space of loop gravity on a fixed graph is the direct product of SU(2) cotangent bundles. The complete phase space of the theory is recovered taking the union over all possible graphs \cite{AAreview}. 

Usually, the new variables $(g_e,X_e)$ are seen as a distributional version of the continuum geometric interpretation. However, one might wonder whether there exists also an interpretation of these variables in terms of \emph{discrete} geometries. In particular, an interpretation which would include the equivalent of the splitting \Ref{Aconn} with a clear separation between intrinsic and extrinsic geometry, and possibly, a nice description of the gauge invariant reduction of the phase space. This is the question we address here, with the viewpoint that this alternative approach gives a new insight on the theory which can turn out to have useful applications.

The idea of labeling the states of loop gravity, at least at the gauge-invariant level, in terms of (some notion of) discrete geometries is not new: indeed long ago Immirzi \cite{Immirzi} suggested a connection with Regge calculus as the analogue for loop gravity of the lattice description of QCD. The connection we find is however not with Regge geometries, but with a looser notion of discrete geometry, which we dub \emph{twisted}.\footnote{The fact that Regge geometries only form a subset of the phase space of loop gravity has already been pointed out by Dittrich and Ryan \cite{BiancaJimmy}. We will come back to this point in Section \ref{SecRegge}.} 
 As we will see, twisted geometries are locally well-defined, but the local patches lack a consistent gluing among each other.
Specifically, unlike Regge calculus where the geometry is given by edge lengths \cite{Regge}, our twisted geometries are described by the assignment to each triangle of its oriented area, two unit normals as seen from the two polyhedra sharing it, and an additional angle related to the extrinsic curvature. Our description holds for both the kinematical and the gauge-invariant phase spaces. The difference between the two is captured by a natural \emph{closure condition} satisfied or not by the labels. Furthermore, we will exhibit in  \cite{Twistor} an intriguing relation between this description and twistors, thus completing the motivations for the name.

The motivation for these labels comes from the coherent intertwiners introduced in \cite{LS} (for earlier ideas, see also \cite{RS}) and further developed in \cite{CF3}, which have proved useful to construct the new EPR-FK-LS spin foam models \cite{LS2,FK}, and crucial to study their semiclassical limit \cite{CF2,CF3,Barrett} and $n$-point correlations \cite{gravEPR}. 
Recall that a standard spin network assigns half-integers $j_e$ to each edge of its graph, and an additional half-integer, called intertwiner, on each vertex. This forms a complete basis in the Hilbert space of loop gravity on a fixed graph, $L^{2}(G_{\Gamma}) = \oplus_{j_e} \left(\otimes_v {\cal H}_{\vec{\jmath}_v} \right)$. The idea introduced in \cite{LS} is to  label each $n$-valent vertex, by $n$ unit vectors in $\R^3$,  instead of the $n-3$ intertwiners. The next key step is the understanding that the labels of the $n$-valent vertex can be restricted to satisfy a certain closure condition (see below) without loosing any information \cite{CF3, Barrett}. This results in $2(n-3)$ labels allowing a geometric interpretation of the states in terms of convex polyhedra \cite{FKL, FLnew}.
With this motivation in mind, we consider the classical system of these labels, namely areas and normals.
What we show in this paper is that these data can be extended to describe the intrinsic \emph{as well as extrinsic geometry of a triangulation} dual to the graph.

\subsection{Area-angle variables}\label{SecAA}

Consider an oriented graph $\Ga$, and for simplicity let us first take it four-valent, so that it is dual to a triangulation. As we will see this restriction can be relaxed without changing our conclusions, but it keep the discussion more focused.
Assign a real number $j_e\in \R$ to each edge, representing the oriented area of the dual triangle, and four unit vectors $N_e(v)\in\R^3$ to each vertex, representing the normals to the four triangles in the dual tetrahedron. 
The whole graph  carries the direct product space 
\begin{align}\label{Pev}
P^{\rm aux}_\Ga \equiv  \BigTimes_{e}\R_{e} \BigTimes_v P_v, \qquad P_v \equiv\BigTimes_{ e:e\supset v}{\cs}^{2}_{e},
\end{align}
where we used the fact that unit vectors define an element on the two-sphere $\cs^2$.
We call this space auxiliary for reasons that will become clear in the following.
Notice that by virtue of our assignment, each edge is labelled by $j_e$ and \emph{two} unit vectors.
Calling  $s$ the source vertex and $t$ the target vertex of an edge $e$, we denote them $N_e=N_e(s)$ and $\tN_{e}=N_{e}(t)$.
We can use this notation to factorize the space as
\begin{align}
\label{Pe}
P^{\rm aux}_\Ga = \BigTimes_e P^{\rm aux}_e, \qquad P^{\rm aux}_e= {\cs}^{2}_e\bigTimes {\cs}^{2}_e \bigTimes \R_e.
\end{align}
The variables associated to each edge of the graph are thus a triple $ (N_e, \tN_{e}, j_e) $. 
Can we use these variables to define a notion of (discrete) metric? Long ago \cite{Regge} Regge showed that a metric on a triangulation is uniquely defined (up to isometries) assigning the edge lengths. Hence, what we need are suitable conditions permitting to reconstruct the edge lengths from our variables.
As it turns out, one needs two types of conditions, called closure and gluing constraints.
The closure constraint is defined on each vertex by
\equ\label{closure}
C_{\vj_v} \equiv \sum_{e\supset v} j_e \, N_e(v) = 0.
\nequ
When the closure is satisfied, the variables $(j_e, N_e(v))$ in the constrained space
\equ
{\rm Tet}_{\vj_v}  =  \left\{N_e(v) \in P_v  \ \big| \ C_{\vj_{v}} = 0 \right\}
\nequ
define the geometry of a flat tetrahedron (for a 4-valent vertex) embedded  in $\R^3$.
The geometry is unique up to rotations, and it is useful to introduce the 
space $S_{\vj_{v}}$ of shapes of a tetrahedron, i.e. the space of closed normals modulo rotations, 
\be\label{Shapes}
S_{\vj_{v}} \equiv {\rm Tet}_{\vj_v}  / \SU(2). 
\ee
At fixed areas, the space is two dimensional and it can be parametrized by two (non-opposite) dihedral angles
$\phi_{ee'}=\arccos(N_e\cdot N_{e'})$  \cite{Barbieri}.
It is also known \cite{Kapovich, CF3} that this space (a) is a symplectic manifold isomorphic to $S^2$, and (b) the SU(2) orbits in Tet$_{\vj_v}$ are generated precisely by \Ref{closure}; hence $S_{\vj_{v}}$ can be obtained imposing the closure constraint and dividing out by the action of the gauge transformation it generates. This symplectic reduction is denoted by the double quotient $S_{\vj_{v}} = P_v/\!/C_{\vj_{v}}$. 
Considering the space \Ref{Pev} on the whole graph, we can apply the symplectic reduction by ${\cal C} = \prod_v C_{\vj_{v}}$, and impose the closure condition on each node. The result is 
\be\label{Pclosed}
{\cal K}_\Gamma \equiv P^{\rm aux}_\Gamma/\!/{\cal C} = \BigTimes_e\R_e \BigTimes_v S_{\vj_{v}}.
\ee
 What is remarkable here is the fact that the closure constraint can be imposed explicitly and locally at each vertex. This result will play an important role later on.

The constrained space ${\cal K}_\Gamma$ of oriented areas and angles defines a precise classical 3d geometry on each tetrahedron. 
The next step is to ensure that the individual tetrahedra glue together to form a consistent geometry on
the whole triangulation. By construction, two neighbouring tetrahedra induce different geometries on the shared triangle, with same area but different shape in general.
To match the shapes one needs additional gluing constraints \cite{BS}, which 
involve only the dihedral angles $\phi_{ee'}$, and are local on each pair of tetrahedra.\footnote{Or equivalent 
constraints written in terms of the normals, derived in  \cite{CF3} and called Regge constraints. Similar constraints appear in \cite{Barrett}, but although they are also called Regge constraints, they are in fact slightly stronger than the ones in \cite{CF3}.}
Together, the closure and gluing constraints guarantee that a unique set of edge lengths, and thus a Regge geometry, can be reconstructed from the areas and angles $(j_e, \phi_{ee'})$ \cite{BS}.

A similar construction can be generalized to vertices of arbitrary valency, but the description of the discrete geometry is now more intricate. As shown in \cite{FKL}, each $n$-valent vertex can be thought of as a flat convex polyhedron with $n$ faces. This polyhedron is such that the polygonal decomposition of its 2d boundary possesses only trivalent vertices, but the boundary faces are not necessarily triangular.
There is a phase space of shapes like \Ref{Shapes} associated to the $n$-vertex, this time $2(n-3)$-dimensional, which is still a symplectic quotient generated by the closure condition, and whose geometrical data can be used to label the quantum states \cite{CF3,FKL,Kapovich}.
Hence, the $n$-valent case is well understood at the level of the closure constraints and the space of shapes of the polyhedron. On the other hand, a deeper analysis is still needed, in particular concerning a suitable generalization of the gluing constraints.

This discussion shows how the reconstruct a discrete metric from the area-angle variables. 
The new question we address in this paper is more ambitious: can we relate the area-angle variables not only to a Regge geometry, but to the full phase space of loop gravity? As we recalled above, also the loop phase space takes a form factorized on edges, thus one can directly look at the edge contributions, respectively $P^{\rm aux}_e$ and $T^*\SU(2)$.
A quick comparison shows immediately that a positive answer to our question is far from obvious: $P^{\rm aux}_e$ has one less dimension, a different topology, and does not seem to carry any information about the connection. If areas and angles are complete geometrical data, we should be able to reconstruct a notion of local frame, and the corresponding rotations mapping one local frame onto the next. Where is the information on such rotation?

\subsection{Twisted geometries}\label{SecTwi}

To address the question, consider two adjacent vertices and the edge connecting them. The five variables $(N_e, \tN_{e}, j_e) $ represent the area of the triangle and its normals in the two frames  
sharing it.
The crucial presence of two normals allows us to write down a natural compatibility condition for a finite connection $g_{e}$
as the group element rotating one normal into the other, i.e.
\equ\label{NRN}
N_e  =R(g_e) \, \tilde{N}_{e}.
\nequ
This equation can be solved for $g_e\in\SU(2)$, with
$R$ the rotation matrix in the adjoint representation.
The connection $g_e$ so introduced defines a notion of parallel transport by which the normal $N_e$
of the triangle in the frame of its ``source'' tetrahedron is mapped into the 
frame of its ``target'' tetrahedron.
This rotation is the answer to our question. 

However, notice that \Ref{NRN} does not \emph{fully} determine $g_e$, 
because it gives only 2 independent equations:
$R(g_e)$ is determined up to rotations along the $\tN$ axis. 
Namely, if $\bar{g}_{e}$ is a solution, then 
$g_{e}(N_e, \tN_{e}, \xi_e) = \bar{g}_{e}\, e^{\xi_e \tilde N^{i}_{e}\tau_{i}}$ is also a solution, for an arbitrary angle $\xi\in[-\pi,\pi]$.
Here and in the following we use the SU(2) fundamental representation, with generators $\tau_i=-i \sigma_i / 2$,
where $\sigma_i$ are the Pauli matrices.
One can easily find a solution to the compatibility condition \Ref{NRN} by 
constructing first a group element $n_{e}$ rotating $\tau_{3}$ into $ N_{e}^i\tau_i$,
that is $R(n_{e}) \tau_{3} = N_{e}^i\tau_i$, and similarly $\tl n_{e}$ for $\tN_e$.
Once $n_{e},\tilde{n}_{e}$ are found, the most general solution of the compatibilty condition is 
 \be\label{gNNt}
 g_{e}(N_e, \tN_{e}, \xi_e) = n_{e}\, e^{\xi_{e}\tau_{3}} \, \tilde{n}_{e}^{-1}.
 \ee
We see that in order to uniquely  define a connection associated to $(N_e$, $\tN_{e})$,
we need an additional angle per edge,  $\xi_e$. 

The above equation is also the key to recover a version of the splitting \Ref{Aconn} on a fixed graph. To that end, consider the projection of \Ref{Aconn} along an infinitesimal edge direction $\ell_e^{a}$, with $A_e\equiv A_a \ell_e^a$, $\Gamma_e\equiv \Gamma_a \ell_e^a$ and $K_e\equiv K_a \ell_e^a$. One can always perform an SU(2) gauge transformation $u$ that diagonalises the extrinsic curvature elements, $ K_{e} = \xi_{e} u \tau_{3}u^{-1}$. 
Then \Ref{Aconn} reads
\be\label{Agf}
A_{e}^{u} = \Gamma^{u}_{e} + \ga \, \xi_{e}\tau_{3},
\ee
where $ \Gamma_{e}^{u} = u^{-1} \Gamma_{e} u + u^{-1}\rd_{e}u$.
We propose to look at the decomposition of $g_{e}$ in \Ref{gNNt} as the discrete  version of \Ref{Agf}, where the gauge-transformed spin connection $\Gamma^{u}$ amounts to the knowledge of $(N_{e},\tilde{N}_{e})$. It follows that in a general gauge, this pair contains information about \emph{both intrinsic and extrinsic} geometry. 
We will come back to this point in Section \ref{SecGeoInt}.

In retrospect, such a splitting also explains our incapacity to interpret purely geometrically the 
data labeled by $(j_{e}, N_{e},\tilde{N}_{e})$, caused by the shape-matching problem raised earlier.
This is because in general these data represent information about both intrinsic and extrinsic geometry,
thus any attempt to interpret them purely in terms of intrinsic geometry is bound to fail.
 Out of this five degrees of freedom three should be purely interpretable as some intrinsic geometrical property 
of the 3d slice, while the other two carry informations about the extrinsic geometry.
The extra angle $\xi_{e}$ is the missing ingredient necessary in order to reconstruct the third component of the extrinsic curvature tensor.\footnote{The same angle and a similar construction were studied by Immirzi \cite{Immirzi}, see more on this in Section \ref{SecSym}. Furthermore, an angle that plays a similar role to $\xi$ is introduced in a covariant context by Dittrich and Ryan \cite{BiancaJimmy} and by Bonzom \cite{Bonzom}. In these works, the four dimensional normals to the tetrahedra are used, instead of the three dimensional normals to the triangles as here. 
}

Including this additional angle, the space of variables associated with an edge of the graph is $6$ dimensional, $(N_e, \tN_e, j_e,\xi_e)$. Accordingly, we define the extended space
\be\label{P}
P_\Ga=\BigTimes_{e} P_e, \qquad P_e = \cs^2_e \bigTimes \cs^2_e \bigTimes T^*S^1_e,
\ee
where we used the obvious isomorphism of ${\mathbbm R}_e \bigTimes S^1_e$ with $T^*S^1_e$, the cotangent space to a circle.
We dub this even-dimensional space of oriented areas, normals and extrinsic geometry the space of 
\emph{twisted} geometries, given the lack of the required constraints to read a Regge geometry off these variables.

The edge component $P_e$ of the space of twisted geometries has on a given edge the same dimensionality of $T^*\SU(2)$.
The first results of this paper are that $P_\Ga$ is a \emph{presymplectic} manifold, and that there exists a reduction $\bar P_\Ga$ of $P_\Ga$ such that\footnote{A manifold is presymplectic if equipped with a closed but possibly degenerate 2-form $\Omega$. The reduction is then the quotient by the kernel of $\Omega$.} 
{
{\renewcommand{\theenumi}{\roman{enumi}}   \renewcommand{\labelenumi}{(\theenumi)}
\begin{enumerate}
\item $\bar P_\Gamma$ is a phase space, in which $(j_{e},\xi_{e})$ are conjugate variables;

\item as a phase space, it is \emph{globally symplectomorphic} to
 the non-gauge-invariant phase space of loop quantum gravity on a fixed graph,  
 $$\bar P_{\Gamma} \cong \BigTimes_{e} T^{*} \SU(2)_{e};$$
\end{enumerate}
Our construction turns out to be very similar to the early attemp by Immirzi \cite{Immirzi}, which our work develops and extends. 

Following the discussion of the previous section, one can also consider the closure constraints on the extended space $\bar P_{\Gamma}$. This gives a notion of \emph{closed} twisted geometries associated to the space
 \be\label{SGamma}
 S_{\Gamma} = \BigTimes_{e} T^{*}S^{1}_e \BigTimes_{v} S_{\vj_{v}},
 \ee
with $S_{\vj_{v}}$ the space of shapes of the polyhedron corresponding to the valency of the vertex.
Our third result is that 
\begin{enumerate}
\setcounter{enumi}{2}
\item 
$S_\Gamma$ is presymplectic, and its reduction $\bar S_{\Gamma}$ is symplectomorphic to the gauge-invariant space of loop quantum gravity,
\be\label{SGamma1}
\bar S_{\Gamma} \cong \BigTimes_{e} T^{*} \SU(2)_{e}/\!/ \SU(2)^{V_{\Gamma}},
\ee
where $V_{\Gamma}$ is the total number of vertices in the graph, and as above the double quotient means 
imposing the Gauss law constraints at each vertex and dividing out the action of the  SU(2) gauge transformation it generates.
\end{enumerate}
}

Twisted geometries are thus particularly useful because they naturally parametrize also the gauge-invariant phase space: the only thing to take into account is the closure condition on the labels, and this can be implemented going from the normals to suitable cross sections \cite{FKL,FLnew}.
The result \Ref{SGamma} implies that the parametrization in terms of twisted geometries factorises the gauge-invariant phase space of gravity as a product of phase spaces associated  with edges and vertices.
This {\it abelianisation} of the gauge-invariant phase space of loop gravity is a remarkable consequence of our parametrization. It provides a classical analogue to the well known factorization of the SU(2) spin network Hilbert space associated with a graph $\Gamma$ as a sum over intertwiner spaces,  $\oplus_{j_e} \left(\otimes_v {\cal H}_{\vec{\jmath}_v} \right)$.
 
 The goal of the next sections 2--4 is to demonstrate  in a precise manner the claims (i), (ii), (iii). This is the bulk of the paper, which means that both the non-gauge-invariant and the gauge-invariant phase spaces of loop gravity can be parametrized in terms of a notion of discrete geometry. These twisted geometries are then candidate labels for full coherent states of the theory.
 These results are the starting point for the construction of new coherent states which will appear in \cite{noiQ}.

 Finally in Section \ref{SecGeoInt}, we will come back to one of the initial questions, how we can separe the intrinsic from the extrinsic metric using the twisted geometries parametrization. We will also comment on the relation between twisted geometries and the construction of a phase space for Regge calculus investigated by Dittrich and Ryan \cite{BiancaJimmy}, in particular comment on $\xi_e$ and the four dimensional dihedral angles.

\section{Phase space of twisted geometries}

Consider the edge space
\equ\label{defP}
P \equiv \cs^2 \bigTimes \cs^2 \bigTimes T^*\cs^1,
\nequ
with variables $(N, \tN, j, \xi)$, where we dropped the label $e$ everywhere to simplify the notation.
Given the Cartesian factorization of $P_\Gamma$ in \Ref{P}, the claims (i) and (ii) follow trivially if we can prove a symplectomorphism to $T^*$SU(2). 
To establish this result, we will proceed in three steps:
\begin{enumerate}
\item We show that there is a natural and in fact unique Poisson  structure on \Ref{defP} which extends the 
Poisson structure of each factor of the Cartesian product.

\item We show that the 2-form $\Omega_P$ associated to the Poisson brackets is degenerate on the codimension one manifold $j=0$. $P$ is thus a \emph{presymplectic} manifold, which turns out to be reducible. The reduction, namely the quotient space in which we divide by the kernel of $\Omega_P$, 
$\bar P \equiv P/{\rm Ker} \, \Omega_P $
is a genuine symplectic manifold, and thus a phase space. 

\item The key result is that there exists an {\it isomorphism} 
which is also a {\it symplectomorphism}  such that
\be\label{main}
\bar P/\Z_{2}\approx {T}^{*}\SU(2),
\ee
where the $\Z_{2}$ identification is given  by $$ \sigma: (N,\tN,j,\xi) \to (-N,-\tN,-j,-\xi).$$ 
This map is a symplectomorphism, therefore the quotient by it is still a symplectic manifold.
 
 \end{enumerate}

Before continuing, let us fix our conventions. Given a symplectic 2 form $\Omega$ we define 
the Hamiltonian vector field $\hat{X}_{f}$ associated with a function $f$ via $i_{\hat X_{f}}\Omega = -\rd f$, and we will often refer to $f$ as the Hamiltonian of $\hat{X}_f$. The Poisson brackets are defined as 
\equ\label{defPP}
\{f,g\} = i_{\hat{X}_{g}}i_{\hat{X}_{f}}\Omega=\Omega(\hat{X}_{f}, \hat{X}_{g})= \hat X_{f}(g).
\nequ

 \subsection{Poisson structure}
Our starting point is that each factor of the Cartesian product \Ref{defP} is a symplectic manifold on its own:
\begin{itemize}
\item 
A cotangent bundle $T^*\cs^1$ has symplectic 2-form $\OmT  
= \rd \xi \w \rd j$, and Poisson bracket
$\{\xi, j\}=1.$

\item
A 2-sphere $\cs^2_R$ of radius $R$ has symplectic 2-form given by the area form 
$\OmS = \pm R \sin \th \, \rd \th \w \rd \phi$, where $\th$ and $\phi$ are the polar and azimuthal angles, and the sign depends on the orientation. The Poisson bracket is conveniently given in terms of the components of the unit vector 
$N=(\cos\phi \sin\th, \, \sin\phi\sin\th, \, \cos\th)$, 
as $\left\{R N^i, R N^j \right\} = \pm \eps^{ij}{}_k R N^k. $ Notice that 
the radius $R$ is a ``Casimir'', i.e. Poisson-commutes with $\theta$ and $\phi$: $\{ N^{i}, R\}=0$. 
\end{itemize}
We extend the above brackets to $P$ taking the two spheres in  \Ref{defP} to have radius $j$ and opposite orientations,\footnote{This is just a choice for later convenience. One could also take the same orientation, as it is done in \cite{Twistor}.} and such that the bracket between $N$ and $\tilde{N}$ vanishes:
This means that we have the following brackets, 
\begin{subequations}\label{PP}\eqa\label{PS2}
&& \{jN^i, jN^j \} = \eps^{ij}{}_k\,j N^k,
\hspace{.8cm} \{j \tl N^i, j \tl N^j \} = -\eps^{ij}{}_k\, j \tl N^k, 
\hspace{.7cm} \{N^i, \tl N^j \} = 0, \\ \nn\\\label{Pzero}
&& \{\xi, j \} = 1, \hspace{2.95cm} \{N^i, j \} = 0, \hspace{2.9cm} \{\tl N^i, j \} = 0.
\neqa
We still have to choose the brackets between $\xi$ and $N,\tilde{N}$. Let us give them in terms of a 
 certain function $L:\cs^2\to \R^3$, 
 \be
 \{\xi, j N^i \} \equiv L^i(N), \hspace{1.6cm} \{\xi, j \tl N^i \} \equiv L^i(\tl N).\label{PTh}
 \ee\end{subequations}
We claim that there is a choice of $L^i$, {\em unique} up to canonical transformations, such that the Poisson algebra closes and $P$ is locally symplectomorphic to $T^*\SU(2)$.
As we will see, the Lie algebra function $L$  is also  uniquely determined by geometric considerations. 

Notice that independently of the relation to $T^*$SU(2), the request that \Ref{PP} closes as an algebra already constrains largely the form of $L^i$. Firstly, we have 
\be
j N_i L^i = j N_i \{\xi, j N^i \} = \f12 \{\xi, j^2 \} = j \{\xi, j \} = j,
\ee
which gives the normalization condition $L \cdot N = 1$. Secondly, from the Jacobi identity 
\be
\{\xi, \{jN^i,jN^j\}\}+\{jN^i, \{jN^j,\xi\}\}+\{jN^j, \{\xi, jN^i\}\} 
\equiv 0
\ee
we get the following coherence identity,
\be\label{PNL}
\{jN^i, L^j(N)\} - \{ jN^j, L^i(N)\} \equiv \eps^{ij}{}_k L^k(N). 
\ee

Similarly, we need $L(\tN) \cdot \tN =1$ and
\be\label{PtlNL}
\{j\tN^i, L^j(\tl N)\} - \{ j\tN^j, L^i(\tl N)\} \equiv -\eps^{ij}{}_k L^k(\tl N), 
\ee
where the minus sign is caused by the reversed orientation in \Ref{PS2}.

The normalizations and coherence identities satisfied by $L$ will turn out to be enough to guarantee the closure of the algebra. Furthermore, we will also prove that these conditions admit a unique solution, up to canonical transformations. These are shifts $\xi\mapsto \xi -\al(N)$, with the other variables unchanged, and induce $L(N)\mapsto L(N)+\{jN,\al(N)\}$. 
In fact, we will see 
that if $L$ and $L'$ are two solutions of $\tr(LN)=1$ and \Ref{PNL}, then there always exists a function $\al(N)$ on the sphere such that 
\be\label{gauge}
{L'}^{i}(N) = L^{i}(N) + \{jN^i, \al(N)\}.
\ee
This shift trivially preserves the normalisation condition,
\equ\label{L'N1}
L'^i N_i = L^{i} N_i + N_i \{jN^i, \al(N)\} = 1 +\f1{2j} \{j^2,\al(N)\} = 1,
\nequ
as well as \Ref{PNL}.
Up to these canonical transformations, there is a unique Poisson structure on $P$ such that its projection on each factor $S^2_j$ and $T^*S^1$ is the canonical one.

To prove this claim, we now construct explicitly the solution $L$. This requires the Hopf map, a standard tool of differential geometry which we  review below.

\subsection{Hopf map}
The Hopf map is a projection $\pi : \cs^3  \to \cs^2 $,
such that every point on $\cs^2$ comes from a circle on $\cs^3$. 
Since SU(2)$\cong \cs^3$ and $S^2\cong \SU(2)/{\rm U}(1)$, where U(1) is the diagonal subgroup generated for instance by $\tau_3$, the map can be defined in terms of group elements. In the fundamental representation of SU(2) as 2-by-2 unitary matrices, the Hopf map reads 
\eqa\nn
\pi : && \SU(2) \to \cs^2 \\ && g \mapsto N(g)=g\tau_3 g^{-1}. \nn
\neqa
The vector $N(g)$ is manifestly invariant under $g \mapsto g^\alpha= g e^{\alpha\tau_3}$, thus it is a function of two variables only.
To be more explicit, we parametrize $g$ with two complex numbers $z_0$ and $z_1$ as follows,
\be
g=\f1{\sqrt{|z_0|^2+|z_1|^2}} 
\left(\barray{cc} \bar z_0 & z_1 \\ -\bar z_1 & z_0 \earray\right).
\ee
A direct calculation then gives
\be \label{HopfSU2}
N(g) = g\tau_3 g^{-1} = \f{1}{1+|z|^2}\big[(1-|z|^2)\tau_3 - z \tau_+ - \bar z \tau_- \big] =
\f{i}{2(1+|z|^2)} \left(\barray{cc} |z|^2-1 & 2z \\ 2 \bar z & 1-|z|^2 \earray\right),
\ee
where $z\equiv z_1/z_0$, and we introduced $\tau_\pm = \tau_1 \pm i \tau_2$. The expert reader will recognize $z\equiv z_1/z_0$ as the Hopf map for the stereographic projection of $S^2$ from the hemisphere with $z_0=0$.

This result shows that SU(2) can be seen as a bundle (the Hopf bundle) over $\cs^2$ with a U(1) fibre. 
On this bundle we can define a section, that is an inverse map 
\eqa\nn
n: && \cs^2\to \SU(2) \\ && N \mapsto n(N)\nn
\neqa
such that $\pi(n(N))\equiv N$, given by
\be\label{nN1}
n(N(z)) \equiv \f1{\sqrt{1+|z|^{2}}} \big[ {\mathbbm 1} +i z \tau_+ - i \bar z \tau_- \big] =
\f1{\sqrt{1+|z|^{2}}}\left( \begin{array}{lr}1 &z \\ -\bar{z} &1 \end{array}\right).
\ee
This section associates an SU(2) element $n$ to each point of the 2-sphere, the latter parametrized by the stereographic projection $z$. For instance taking the projection from the south pole, we have $z=-\tan \f\th2 \, e^{-i \phi}$ in terms of the familiar polar coordinates $(\th,\phi)$. The choice \Ref{nN1} is clearly not unique, and any $n^\alpha \equiv n e^{\alpha\tau_3}$ defines an alternative section.
In the rest of the paper, we will work extensively with the section $n(N(z))$ defined by \Ref{nN1}. 
With abuse of notation, we will use simply $n$
to refer to it, and we will interchangeably consider $n$ as a function of $N$ or as a function of 
$N(z)$, hence $z$,  even if we do not explicitely write the argument of $n$. 
We will also write the Hopf map simply as $N=n\tau_3 n^{-1}$.

\subsection{Geometric action on the Hopf section}\label{secGeoA}

Let us fix some conventions about SU(2).
The $\mathfrak{su(2)}$ algebra generators are
$\tau_{i} = -i \sigma_{i}/2$ and satisfy $[\tau_{i}, \tau_{j}] = \epsilon_{ijk} \, \tau^{k}.$
We introduce a cyclic trace $\tr(XY) \equiv -2 \, \mathrm{tr}_{1/2}(XY)$,
for any $(X,Y) \in \mathfrak{su(2)}$, where $\mathrm{tr_{1/2}}$ is the trace in the 
fundamental representation. 
Tr gives a positive pairing, invariant under adjoint action, such that $\tr(\tau_{i}\tau_{j}) = \d_{ij}$.
We can then identify $\mathfrak{su(2)}$ with $\R^{3}$ via $X^i\equiv \tr(X\tau^i)$, and $\tr(XY)=X_i Y^i$. 

Next, an element $N\in S^2$ is a unit vector in $\R^3$, and we have a natural action of the group SU(2) by rotations.
This action is given, once we represent the 2-sphere as embedded into the  Lie algebra $\mathfrak{su(2)}$, via the coadjoint representation. Hence we can associate 
to an algebra element $X\in \mathfrak{su(2)}$ a vector field $\hat X$ on $\cs^2$, and its action on functions on the sphere is given by
\be\label{LXf}
\LX f(N) = \f{d}{d t} f\left(e^{-t X} N e^{t X}\right)\Big|_{t=0}.
\ee
where $\LX\equiv i_{\hat X}\rd + \rd i_{\hat X}$ denotes the Lie derivative and $i$ the interior product.
For linear functions we have
\be\label{LXN}
{\cal L}_{\hat X}N = -[X,N].
\ee
Viewing $S^{2}$ as a symplectic manifold, the action of SU(2) on it is Hamiltonian, i.e. by explicit calculation one can verify that $\hat X$ is an Hamiltonian vector field -- associated to the function $h_{X}(N)\equiv j \tr (NX)$, and the above Lie derivative is obtained from the Poisson bracket between $N$ and $h_{X}$,
\be\label{Hamaction}
\{h_{X}, N\} = \OmS (\hat X, \hat {N} ) = - [X,N] = \LX N.
\ee
This Hamiltonian action can be used to write the coherence identity \Ref{PNL} as an identity involving Lie derivatives:
contracting (\ref{PNL}) with $X_{i}$ and $Y_{j}$, we get
\be
{\cal L}_{\hat{X}}L_{Y} -{\cal L}_{\hat{Y}} L_{X} =  L_{[X,Y]},
\ee
where $L_{X}\equiv \tr(LX)$ is the component of $L$ along the algebra element $X$.

We are now interested in the action of the algebra on the Hopf section \Ref{nN1}. That is, 
we want to define the Lie derivative on $n(N)$, in such a way that the section is preserved. 
Let us first notice that 
\be
\LX N(n) = (\LX n) \tau_3 n^{-1} + n \tau_3 (\LX n^{-1})= [(\LX n)n^{-1}, N(n)].
\ee
Comparing this to \Ref{LXN}, we deduce that 
\be\label{defF}
(\LX n)n^{-1} = - X + N(n) F_{X}(N),
\ee
where $F_{X}(N)$ is a function on the sphere and $N(n) F_{X}(N)$ commutes with the algebra element $N(n)$. 
We see that $F$ acts as a connection in preserving the Hopf map section under Lie derivative. Remarkably, \emph{this connection turns out to satisfy precisely the normalization and coherence identity required for the above function $L$}. Therefore we can identify $L$ as the section-preserving connection, which gives it a clear geometrical meaning. 
This key result is established thanks to the following lemma.

\medskip
{\bf Lemma}.

\emph{There is a unique function $L: \cs^{2}\to  \mathfrak{su(2)}$ that we call the ``Lagrangian'', such that 
\equ\label{Ldef}
\tr(L\,\rd n n^{-1}) =0 \quad\mathrm{and} \quad
\tr(L N ) = 1,
\nequ
explicitly given by
\be\label{Lz}
L(z) = \tau_3 - \f{z}{2} \tau_+ - \f{\bar z}{2} \tau_- \, .
\ee
\indent The Lagrangian appears in the Lie derivative of the Hopf map section $n(N)$, 
\be\label{Lder}
\LX n = \big( - X + N L_{X}\big) n,
\ee
and it satisfies the key coherence identity
\be\label{Lcoherence}
{\cal L}_{\hat{X}}L_{Y} -{\cal L}_{\hat{Y}} L_{X} =  L_{[X,Y]}.
\ee
Finally, the general solution to this identity satisfying the normalisation condition is given by
\be 
\label{LLdt} L'= L + \rd \al
\ee
where $\al$ is an arbitrary function on the sphere.\footnote{Here \Ref{LLdt} is just a coordinate-independent version of \Ref{gauge}.}
}

\medskip

{\bf Proof.}
The first part of the lemma amounts to simply solving the system \Ref{Ldef}. We can do so using the definition \Ref{HopfSU2} of $N(n)$, and computing the right-invariant 1-form of the section \Ref{nN1},
\be\label{dnn}
\rd n(z) n(z)^{-1} = \frac{i}{1+|z|^{2}} \big\{(\bz \rd z -z\rd \bz) \tau_{3} + \rd z \, \tau_{+} - \rd \bar{z} \, \tau_{-}\big\}.
\ee
Explicitly evaluating \Ref{Ldef} leads to the unique solution \Ref{Lz}.

To prove the second part of the lemma, we take the interior product of an arbitrary vector field $\hat{X}$ with the defining expression \Ref{Ldef}. Recalling that by definition of Lie derivative, 
$(\LX n )n^{-1}=i_{\hat{X}} (\rd n n^{-1})$, we have
\be\label{ia}
0= i_{\hat{X}}\tr(L\,\rd n n^{-1}) = \tr(L \left({\cal L}_{\hat{X}}n \right) n^{-1} ) =
-\tr(L X) + F_{X} \tr(LN)  = - L_{X}  +F_{X},
\ee
where we used \Ref{defF} and the normalization condition. Hence $F_{X}=L_X$ and (\ref{Lder}) is obtained. 

Finally, to prove \Ref{Lcoherence} we first observe that
 \eqa
\LX \left(\rd n n^{-1} \right) &=& i_{\hat X}(\rd n n^{-1} \w \rd n n^{-1} )+ \rd[(\LX n) n^{-1}]            
 \nn\\ &=& [-X+NL_X, \rd n n^{-1}]+ \rd (-X + N L_X) 
\nn\\ &=& N \rd L_{X}  - [X,\rd n n^{-1}], 
\neqa
where we used the definition of Lie derivative in the first equality,
(\ref{Lder}) in the second and $\rd N= [\rd nn^{-1}, N]$ in the third. Therefore, 
\be
0= {\cal L}_{\hat{X}}\tr(L\, \rd n n^{-1}) = \tr\left( ({\cal L}_{\hat{X}}L - [L,X]) \rd n n^{-1}\right)  + \rd L_{X}\,.     \ee
Then, by taking the interior product of this relation with $\hat{Y}$ we get 
 \eqa
 {\cal L}_{\hat{Y}} L_{X} &=& \tr\left( ({\cal L}_{\hat{X}}L - [L,X])( Y - N L_{Y})\right)  \nn \\ \nn   
 &=&{\cal L}_{\hat{X}}L_{Y} - L_{[X,Y]} + L_{Y}\left\{\tr({\cal L}_{\hat{X}}L N)  - \tr (L [X,N]) \right\}\\      
 &=&{\cal L}_{\hat{X}}L_{Y} - L_{[X,Y]} + L_{Y} {\cal L}_{\hat{X}}\tr(LN), 
 \neqa
 and since the last term vanishes, we obtain the coherence identity (\ref{Lcoherence}).
 
 Notice that the latter also ensures that \Ref{Lder} verifies the consistency condition for Lie derivatives, namely ${\LX}{\cal L}_{\hat Y}-{\cal L}_{\hat Y}{\LX}={\cal L}_{[\hat X, \hat Y]}$.
 
Let us now suppose to have another solution $L'$ to the coherence identity and the normalisation condition $\tr({L}'N)=1$. If we define the 1-form $\beta \equiv - \tr({L}'\, \rd n n^{-1})$, we see that
 \be
 \beta_{X}\equiv i_{\hat X}\beta = - \tr(L' (\LX n) n^{-1}) = L'_{X} - {L}_{X}.
 \ee
is a solution of the coherence identity as a sum of two such solutions.
This, together with the definition of the differential 
$i_{\hat{X}}i_{\hat Y} \rd \beta = {\cal L}_{\hat Y} \beta_{X}-\LX \beta_{Y}  +\beta_{[X,Y]}$, implies that $\rd\beta=0$.
Since $H_{1}(S^{2})$, the first homology group of the sphere, is trivial this means that there exist a function $\al$ such that 
$\beta =\rd \al$, and thus ${L}'_{X}=L_{X} + \LX \al$. Finally, $\tr( L' N ) =1 $ follows as in \Ref{L'N1}.
This proves the gauge freedom \Ref{LLdt}, which can be written as anticipated above in \Ref{gauge} using the  Hamiltonian action \Ref{Hamaction}. 
$\square$

\medskip

The gauge freedom \Ref{LLdt} has a very simple geometric origin: it corresponds to a change of section $n\mapsto n^\al \equiv n e^{\al\tau_3}$ with $\al=\al(N)$ a function on the sphere. Indeed, 
\be
(\LX n^\al)(n^\al){}^{-1} = (\LX n)n^{-1} + N \LX\al =- X + N (L_X+\LX\al),
\ee
namely ${L}^\al_{X}=L_{X} + \LX \al$ is the Lagrangian preserving the new section $n^\al$.\footnote{An interesting example (see \cite{Twistor}) is 
$\al(N(z)) = - 2 \, {\rm arg} (z)$, which gives in components $$
\tl L^i(z) = L^i(z) - \{jN^i, 2 \, {\rm arg} (z)\} = (-z^{-1} , -\bar z^{-1} , -1) \equiv - L^i\big(-1/\bar z\big).
$$}

The same Lagrangian \Ref{Lz} can be taken also for the sphere with negative orientation, as we did in \Ref{PTh}. This is because the reversed orientation can be obtained with the map $n\mapsto n\eps$, $\eps=i\sigma_2$, which does not affect the first condition in \Ref{Ldef}. Hence the reversed orientation simply amounts to $N\mapsto -N$, $L\mapsto -L$ and the sign of the bracket is preserved.

From now on, we use the notation $L$ to refer exclusively to \Ref{Lz}. 
The above lemma proves that the Lagrangian $L^i=\tr(L\tau^i)$, i.e. the section-preserving connection,\footnote{
The name ``Lagrangian'' comes from the fact that $L$ is the generator of vertical shifts in the bundle $\cs^2\times U(1)$.
That is under a finite rotation the section $n$ is corrected by the ``action'' $\int L$:
\be
n(e^{-X}Ne^{X}) = e^{-X} n(N) e^{\tau_{3}\int_{0}^{1} L_{X}(e^{-uX}Ne^{uX})\rd u}.
\ee
This formula is the integral version of \Ref{Lder}.
}
satisfies the normalisation condition $L^iN_i=1$ and coherence identities (\ref{PNL}) and (\ref{PtlNL}). Concerning the latter, the minus sign on the right hand side of \Ref{PtlNL} is included automatically in the right hand side of \Ref{Lcoherence}: as we change the orientation of the sphere, we also change the sign of the commutator $[X,Y]$.
These are the necessary conditions to the closure of the algebra discussed earlier. To complete the proof that the Poisson algebra \Ref{PP} closes, we now exhibit the symplectic 2-form, and discuss its properties.

\subsection{Symplectic Potential}

It is convenient for the following to introduce the symplectic potential $\Theta$, or canonical 1-form, such that $\Omega = -\rd \Theta$.
In the case of the cotangent bundle to the circle, the symplectic potential is $j \rd \xi$, and the Hamiltonian vector fields are $\hat \jmath = -\p_\xi$,  $\hat \xi = \p_j$. In the case of a 2-sphere of radius $j$, say with right-handed orientation, we can use the Hopf section to write the potential as
\be
\ThS(N) = j \tr(N \rd n n^{-1})  
= j (\cos\th-1) \, \rd \phi,
\ee
as one can easily verify using \Ref{HopfSU2} and \Ref{dnn}. The associated 2-form reads
\be
\OmS = - \rd \ThS = j \tr(N \rd n n^{-1} \w \rd n n^{-1})  
= j \sin\th \, \rd \th \w \rd \phi,
\ee
and as already discussed in Section \ref{secGeoA}, the Hamiltonian vector fields are the $\hat X$'s generating the adjoint action \Ref{LXN}, and associated to $X\in\mathfrak{su(2)}$. Indeed, using our section-preserving Lie derivative \Ref{Lder}, we can verify that
\be\label{OS}
i_{\hat X} \OmS =  j \tr(N [(\LX n)n^{-1}, \rd nn^{-1}])
= j \tr(X [N, \rd nn^{-1}])=  
- j \rd N_{X},
\ee 
where $N_{X}\equiv \tr(NX)$.
In other words, $\hat X$ is an Hamiltonian vector field with Hamiltonian the linear 
function $h_{X}(N)= jN_{X}$ on the sphere.

Let us now come to the Poisson algebra \Ref{PP} of the twisted geometries, and discuss its symplectic 2-form $\Omega_P=-\rd\Theta_P$.
By inspection, we see from the brackets \Ref{PS2} and \Ref{Pzero} that $\Omega_P$ should contain $\OmT$
and two $\OmS$ with radius $j$ and opposite orientations. 
Remarkably, $\Theta_P$ turns out to be simply a sum of the elementary symplectic potentials of each factor in the cartesian decomposition $P = S^{2} \times S^{2} \times T^{*}S^{1}$, 
\eqa\label{ThetaP}
\Theta_{P} &\equiv& \ThS(N) + \ThS(\tN) + \ThT(j,\xi)
\nn \\ &=& j \tr(N \rd n n^{-1}) - j \tr(\tN \rd \tl n \tl n^{-1}) + j \rd\xi.
\neqa
That is, we claim that the brackets \Ref{PP} are given by 
\bea\label{OmegaP}
\Omega_{P} = -\rd \Theta_P &=& j \tr(N \rd n n^{-1} \w \rd n n^{-1}) - j \tr(\tN \rd \tl n \tl n^{-1} \w \rd \tl n \tl n^{-1}) +
\nn\\ && -\rd j \w \Big[ \rd \xi + \tr(N \rd n n^{-1}) - \tr(\tN \rd \tl n \tl n^{-1})\Big].
\eea

To show it, we first need to compute the Hamiltonian vector fields on $P$. To avoid confusion with the vector fields on $S^2$ and $T^*S^1$ previously introduced, we denote the new ones $\chi_{f}$, where $f = j, \xi, h_X \equiv jN_{X}$ and $\tilde{h}_{X}\equiv j\tilde{N}_{X}$, and such that $i_{\chi_{f}}\Omega_P = -\rd f$.
For $j\neq 0$, these are given by
\begin{subequations}\label{Hvf}\begin{align}\label{Hvf2}
& \chi_{h_{X}} = \hat X - L_X(N) \p_\xi, && \chi_{\tilde{h}_{X}}= -\hat{\tilde{X}} - L_X(\tilde{N}) \p_\xi, \\
\label{Hvf1}
& \chi_{\jmath} = -\p_\xi, && \chi_{\xi} = \p_j + \f1j \widehat{[L,N]} + \f1j \widehat{[L,\tN]}.
\end{align}\end{subequations}
Here $\hat X$ and $\widehat{[L,N]}$  are the vector fields \Ref{LXf} generating the adjoint action on the sphere labelled by $N$, associated respectively with the algebra elements $X$ and $[L(N),N]$.
Similarly, $\hat{\tilde{X}}$ and $\widehat{[L,\tN]}$ are the vector fields \Ref{LXf} generating the adjoint action on the sphere labelled by $\tilde{N}$, associated respectively with the algebra elements $X$ and $[L(\tilde{N}),N]$.

{\bf Proof. } To check \Ref{Hvf2}, we first notice that for a constant $X$ we have 
\be\label{OmY}
i_{\hat{X}} \Omega_{P} =  - \rd \left(j N_{X}\right) + L_{X}(N) \rd j.
\ee
Since $i_{\p_\xi}\Omega_P = \rd j$, \Ref{Hvf2} follows.
The computation for $\chi_{\tilde{h}_{X}}$ is similar up to a sign due to the reversal of the orientation.
To check $\chi_\xi$, we first evaluate
\be
i_{\p_j}\Omega_P = -\rd \xi - \tr (N\rd nn^{-1}) + \tr (\tN\rd \tl n \tl n^{-1}).
\ee
Then,
 \bea
  i_{\widehat{[L,N]}} \, \Omega_{P} &=& 
  j \tr\left( [N,[L,N]]\, \rd {n} {n}^{-1}\right) + \rd j \tr( (N-L)[L,N]) \nn \\ &=& \label{OmLN}
  j \tr\left( (\tr(LN) N - L) \, \rd{n}{n}^{-1}\right) = j \tr\left(  N \, \rd{n}{n}^{-1}\right)\,,
 \eea
where we have used again \Ref{Lder} and in the last equality the defining property of the Lagrangian.
A similar calculation shows that
 \be
  i_{\widehat{[{L},\tilde{N}]}}\, \Omega_{P}
 =  - j \tr( \tilde{N}\, \rd \tilde{n} \tilde{n}^{-1}),
 \ee
thus \Ref{Hvf1} follows. $\square$

Comparing the Hamiltonian vector fields \Ref{Hvf} to those of $T^*S^1$ or $S^2$ on their own, we see that the only differences are the pieces depending on the Lagrangian $L$, which are generated by the last two terms of \Ref{OmegaP}, the ones mixing the circle and the spheres. From the viewpoint of the symplectic potential \Ref{ThetaP}, the only origin of the mixing is the choice of radius of the sphere.\footnote{This almost-diagonal form of the symplectic potential has a deeper origin which will be investigated in further studies, and it is crucial for the quantization of this phase space.}
Using these vector fields and properties such as (\ref{OmY}, \ref{OmLN}), it is straightforward to check that we recover the Poisson brackets \Ref{PP}, with $L^i = \tr(L\tau^i)$. 
For instance,
 \be
\{\xi, j\} = \Omega_P(\chi_\xi,\chi_j) = 1, \quad \quad 
\{\xi,  j{N}^{i}\} =  \chi_{jN^{i}}(\xi) = -\chi_{\xi}(jN^{i})= L^{i}(N). 
\ee

The Poisson structure \Ref{PP} that we introduced on $P$ can be now studied through the symplectic 2-form \Ref{OmegaP}.
In particular, the Jacobi identity and closure of the algebra can be proved deriving the Maurer-Cartan equation satisfied by this symplectic potential. We leave this demonstration to the interested reader. More important is the issue of the invertibility of $\Omega_P$, because to be a proper symplectic manifold, a space needs to be
equipped with a closed and \emph{non-degenerate} 2-form. Now, $\Omega_P$ is trivally closed, as it descends from a local symplectic potential. However, it is degenerate at $j=0$. Since $\Omega_P$ fails to be non-degenerate everywhere, $P$ is only a \emph{presymplectic} manifold, in the terminology of Souriau \cite{Souriau}. 

A symplectic manifold can then be obtained in two ways. The first way is to simply consider a new space $P^*$ with $j\neq 0$. The second is to reduce the presymplectic manifold by the kernel of $\Omega_P$, i.e. to consider the quotient manifold $\bar{P}\equiv P/ \mathrm{Ker}(\Omega_{P}) $. The latter is a symplectic manifold, with non-degenerate 2-form given by $\Omega_P$  projected on $\bar P$.
It is this second way that leads to a symplectic manifold globally isomorphic to $T^*\SU(2)$, and that we now describe in detail.

Taking the quotient means dividing by the equivalence class $p\sim p'$ where $p'= e^{\hat D} p $, with $\hat{D} \in \mathrm{Ker}(\Omega_P)$ and $p, p' \in P$.
The first thing to do is thus to identify the vectors generating the kernel of $\Omega_P$. As seen, the kernel has support at $j=0$, thus we are looking for the vector fields whose interior product with $\Omega_{P}$ is proportional to $j$, and which leave us on this submanifold. The set of such vector fields is given  by
\be\label{nullvec}
 \hat{D}_{X}\equiv  \chi_{h_{X}} - \chi_{{\tilde h}_{Y}} , \quad \mathrm{with} \quad Y= g^{-1}Xg,
 \ee
where $g=n e^{\xi\tau_{3}}\tilde{n}^{-1}$ is the group element rotating $N$ into $\tN = g^{-1}Ng$. Indeed, using the fact that $N_{X}=\tN_{Y}$, the interior product with the symplectic 2-form computes to
\bea\label{laurent}
i_{\hat{D}_{X}}\Omega_{P} &=& -\rd \left(jN_{X}-j\tilde{N}_{Y}\right) - j\tr(\tilde{N} \rd Y)\\
&=& - j \tr( [N,X] \rd g g^{-1}),
\eea
which vanishes at $j=0$.

Next, to find the equivalence class generated by the vector fields, we notice that they rotate jointly  
the vectors $N $ and $\tN$: $\hat{D}_{X}(N) = -[X,N]$, $\hat{D}_{X}(\tN) = -g^{-1}[X,N] g$. 
Furthermore, this rotation is such that it  preserves the group element $g=n e^{\xi\tau_{3}}\tilde{n}^{-1}$, since
\be
\hat{D}_{X} g = -X g + gY =0.
\ee
Therefore the equivalence relation between $p\in P$ and $p'= e^{\hat D} p \in P$ is $(N,\tN,0,\xi)\sim (N',\tN',0,\xi')$ with $N$ and $\tN$ jointly rotated, and $\xi$ translated in such a way as to preserve $g=n e^{\xi\tau_3} \tilde n^{-1}$. 
Dividing by this equivalence relation means that the two 2-spheres sitting at the boundary are identified, and one is left with a 3-sphere parametrized by $N$ and $\xi$. 

We conclude that the coordinates of 
$
\bar P \equiv P/{\rm Ker} \, \Omega_P 
$
span $P^* = \cs^2_j \bigTimes \cs^2_j \bigTimes T^*\cs^1$ for $j\neq 0$, and $S^3$ for $j=0$.
The two symplectic manifolds $P^*$ and $\bar P$ clearly differ in their topology. The difference is captured by the homotopy group, which is $\mathrm{U}(1)$ for $P^{*}$, and trivial for $\bar P$.\footnote{This difference has implications at the quantum level \cite{noiQ}, where it implies that there is an ambiguity in the quantisation of $P^{*}$, labelled by an angle in $\mathrm{U}(1)$ \cite{Woodhouse}, while the quantisation of the reduced space $\bar P$ is unique and amounts to fix this angle to $0$.
}

Finally, let us point out that the symplectic potential is invariant under the $\Z_{2}$ transformation 
\be\label{Z2}
(N,\tilde{N}, j, \xi)\to (-N,-\tilde{N}, -j, -\xi).
\ee 
This is implemented on the Hopf section by $n\to n\epsilon$ and $\tilde{n}\to \tilde{n}\epsilon$, with $\epsilon=i\sigma_{2}$. 
It is easy to see that this transformation leaves $\Theta_P$ invariant since $\rd(n\epsilon) (n\epsilon)^{-1}=\rd n n^{-1}$. Hence \Ref{Z2} is a canonical transformation, and both $P^*/\Z_{2}$ and $\bar P/\Z_{2}$ are still symplectic manifolds.

\medskip
This concludes the construction of the phase space of twisted geometries. We have shown that $P^*$ and 
$\bar P$, as well as their reductions by \Ref{Z2}, are symplectic manifolds. Their closed, non-degenerate symplectic 2-forms are given by the relevant projections of $\Omega_P$. Specifically, $\bar P$ is given by $P^*$ completed by a 3-sphere sitting at $j=0$. In the next Section, we prove that $\bar P/\Z_2$ is nothing new:
it is isomorphic to the standard phase space used in loop quantum gravity.

\section{Symplectomorphism with SU(2) phase space}

A Lie group $G$ is a manifold, and as for any manifold, a symplectic structure can be associated to
its cotangent bundle, $T^*G$. Let us recall the basics of this construction, refering the reader to
the literature \cite{Arnold,Alekseev} for the details. 

\subsection{$\SU(2)$ cotangent bundle}
The Lie algebra $\mathfrak{g}\cong T_e G$ is isomorphic to the set of right-invariant
vector fields on $G$.\footnote{The group acts on itself by either left or right multiplication. 
Both actions can be used to get an isomorphism of vector fields with the Lie algebra, and to trivialize
the cotangent bundle. Here we choose the right multiplication, but a similar construction 
can be carried over choosing the left multiplication.
} 
A right-invariant vector field in the direction of $X\in\mathfrak{g}$ , which we denote $\n^{\scr L}_{\scr X}$, 
acts on functions on the group as the \emph{left} derivative 
\be\label{leftDer}
 \n^{\scr L}_{\scr X} f(g) \equiv 
\f{\rd}{\rd t} f( e^{-t X} g )\Big|_{t=0},
\ee
Under the adjoint transformation $g\mapsto g X g^{-1}$, we obtain the right derivative
\be\label{righttDer}
\n^{\scr R}_{\scr  X} f(g) \equiv 
\f{\rd}{\rd t} f( g e^{t X} )\Big|_{t=0} = -\n^{\scr L}_{({\scr  g  X g^{-1}})} f(g).
\ee
The map from the vector fields to elements $X$ of the algebra is provided by the algebra-valued,
right-invariant 1-form $\rd g g^{-1}$, which satisfies
\be
i_{\hat X}(\rd g g^{-1}) = (\LX g) g^{-1} = - X.
\ee
The set of right-invariant 1-forms is isomorphic to the dual algebra $\mathfrak{g}^*$, thus the cotangent bundle
trivializes as $T^*G=G\times \mathfrak{g}^*$. 

To study functions on $T^*G$, recall that each element $X\in \mathfrak{g}$ determines a linear function $h_X$ on the dual algebra $\mathfrak{g}^*$. Let us fix from now on $G=\SU(2)$, and take the trace $\tr(XY) \equiv -2 \, \mathrm{tr}_{1/2}(XY)$ introduced earlier. With this ad-invariant positive pairing we can define
the linear action as $h_X(Y^*) = \tr(XY)$ and identify $\mathfrak{su(2)}$ and its dual $\mathfrak{su(2)}^{*}$.
Thanks to this identification we can parametrize $\mathfrak{su(2)}^{*}$ directly with elements $X$ of the algebra.
The 6-dimensional cotangent bundle $T^*G$ is then trivialized by the following symplectic potential \cite{Arnold, Alekseev},
\eqa\nn
 G\times \mathfrak{g}^* &\to& T^{*}G\\
(g,X) &\mapsto& \Theta = \tr(X \rd g g^{-1}). \label{TG}
\neqa
The symplectic 2-form computes to
\be
\Omega = - \rd \tr \left( X \rd g g^{-1}\right)
=\frac12 \tr\left( \rd\tilde{X} \wedge g^{-1} \rd g - \rd X \wedge  \rd g g^{-1} \right)
\ee
where we have introduced $\tilde{X} \equiv -g^{-1}Xg$.
From the symplectic 2-form one gets the following Poisson brackets, 
\begin{align}\label{PT*G}
& \{h_Y, h_Z\} = h_{[Y,Z]}, 
& \{h_Y, f(g) \} = \n^{\scr L}_{\scr Y} f(g),
&& \{f(g), h(g) \} = 0.
\end{align}

{\bf Proof.} Let us identify $ \mathfrak{su(2)}$ with $\mathbbm{R}^{3}$, via $X^i= \tr(\tau^i X)  = h_{\tau^i}(X)$.
Consider then the following vector field on $T^{*}G$,
\be\label{hY}
\hat{Y} \equiv \nabla_{Y}^{L} + [X,Y]^{i} \frac{\partial}{\partial X^{i}}.
\ee
This vector field is such that
\begin{align}
& i_{\hat Y}\Theta = - \tr(XY),
\\
& {\cal L}_{\hat Y}\Theta = \tr([X,Y]\rd g g^{-1}]) - \tr(X[Y,\rd g g^{-1}]) = 0.
\end{align}
Therefore
\be
i_{\hat Y}\Omega = \rd i_{\hat Y}\Theta - {\cal L}_{\hat Y}\Theta = -\rd \tr(XY),
\ee
that is \Ref{hY} is the Hamiltonian vector field of $h_Y(X)$,	
and
$$\{h_{Y},h_{Z} \} = \Omega(\hat Y, \hat Z) = - i_{\hat{Z}} \rd h_{Y} = h_{[Y,Z]} . $$

Next, the Hamiltonian vector field of a function $f(g)$ on the group is
\be
\hat f = - \n_{i}^{\scr L}f \f{\p}{\p X^{i}},
\ee 
since
$$
i_{\hat{f}}\Omega =  \sum_{i} \n_{i}^{\scr L}f \tr\left(\tau^{i}  \rd g  g^{-1} \right) \equiv -\rd f. $$ 
It is then easy to see that two functions on the group, say $f(g)$ and $h(g)$, have a vanishing Poisson bracket,
$\Omega_{T^{*}G}(\hat{X}_{f}, \hat{X}_{h}) =0$. Finally, 
\be
\{h_{Y}, f\}= - i_{\hat Y} \rd f = i_{\hat X_{f}} \rd h_{Y}= \nabla_{\scr Y}^{\scr L}f.
\ee
$\square$

We see from the brackets \Ref{PT*G} that the Poisson action of $h_{Y}(X)$ generates left derivatives. Similarly, the right derivative $\{\tilde{h}_X, f(g) \} = \nabla_{\scr Y}^{\scr R} f(g)$ is generated by the action of $ \tilde{h}_{Y}(X) \equiv \tr(Y \tilde{X})$. Finally, the two Hamiltonians commute: $\{h_{Y},\tilde{h}_{Z}\}=0$.

\subsection{Symplectomorphism}\label{SecSym}

The key to construct the isomorphism is the Hopf map \Ref{HopfSU2} introduced above.
We consider the two sections $n(N)$ and $\tl n(\tN)$ such that
$ N = n \tau_3 n^{-1},$ $\tN = \tl n \tau_3 \tl n^{-1}.$
Then, we define the map 
\begin{subequations}\label{map}\eqa\label{mapX}
(N, \tN, j, \xi) \rightarrow (X, g) \ : \qquad
X &=& j n \tau_3 n^{-1} \\ \label{mapg} 
g &=& n e^{\xi \tau_3} \tl n^{-1}
\neqa\end{subequations}
which implies that $\tilde X \equiv -g^{-1} X g = -j \tN$.
The map is two-to-one, as the two configurations $(N,\tN,j,\xi)$ and $(-N,-\tN,-j,-\xi)$ give the same pair $(X,g)$,
and it can be inverted in each branch provided $|X|\neq 0$: 
\be
j = \pm |X|, \qquad N = \pm \f{X}{|X|}, \qquad \tN = \pm \f{g^{-1}Xg}{|X|}\qquad \xi = \pm \tr(\tau_3 \log(n^{-1} g \tl n)).
\ee
The map then gives an isomorphism
\begin{subequations}\label{iso}\be\label{iso1}
P^*/\Z^2 \cong T^*\SU(2) \setminus \{|X|=0\},
\ee 
where the $\Z^2$ symmetry is the identification \Ref{Z2} of the two configurations with opposite signs all over.
Furthermore, the isomorphism extends trivially to an isomorphism 
\be\label{iso2}
\bar P / \Z^2 \cong T^*\SU(2),
\ee \end{subequations}
since we have already shown that the two spaces coincides at the origins $j=0$, $X=0$, where both are given by a 3-sphere.

What we want to prove next is that the isomorphism \Ref{iso} is also a symplectomorphism, namely it preserves the Poisson structure of the symplectic spaces. As we now show, this is a direct consequence of the identification of the two symplectic potentials \Ref{ThetaP} and \Ref{TG}.

\medskip
{\bf Proposition.} \emph{The map \Ref{map} provides an invertible symplectomorphism between the phase space
$\bar P_e$ with Poisson brackets \Ref{PP} and $T^*\SU(2)$ with Poisson brackets \Ref{PT*G}.}

\medskip
{\bf Proof.} A straighforward calculation gives
 \bea\label{ThetaTheta}
 \Theta_{T^{*}G}=\tr({X}\, \rd g g^{-1} ) 
 &=&  j \tr\left(n \tau_{3}n^{-1} \, \left( \rd {n} {n}^{-1}+  {n} \rd \xi  \tau_{3}  {n}^{-1}
 - n e^{\xi\tau_{3}}\tilde{n}^{-1} \rd \tilde{n} \tilde{n}^{-1} e^{-\xi \tau_{3}}n^{-1} \right) \right)  \nonumber \\
 &=& j  \tr\left(N \, \rd {n} {n}^{-1}\right) + j\rd \xi - j  \tr\left(\tilde{N} \, \rd \tilde{n} \tilde{n}^{-1}\right) 
= \Theta_{P}.
 \eea
The identification of the two potentials is up to the $\Z_2$ symmetry \Ref{Z2}, which as already discussed leaves $\Theta_P$ invariant. $\square$

Let us make some remarks.

\begin{itemize}

\item Even if the proof is straighforward, it is instructive to check the symplectomorphism at the level of Poisson brackets.
To that end, it is convenient to use the identification of $\mathfrak{su(2)}$ with $\mathbbm{R}^{3}$ via $X^i= \tr(\tau^i X)$, and write the Poisson brackets \Ref{PT*G} of linear functions on $T^*\SU(2)$ in the simple form
\begin{align}\label{PT*G1}
& \{X^i, X^j \} = \eps^{ij}{}_k X^k, 
& \{X^i, g \} = -\tau^i g,
&& \{\tilde X^i, g \} =  g \tau^i.
\end{align}
The first Poisson brackets in \Ref{PT*G1} can be immediately
verified using the Poisson brackets \Ref{PS2} in the definitions \Ref{mapX}.
The second Poisson bracket can be verified as follows.
Using \Ref{mapX} and \Ref{mapg}, we have
\be\label{Xg1}
\{X^i, g\} = \{ jN^i, n e^{\xi \tau_3} \tl n^{-1}\} = \{jN^i, n \} e^{\xi \tau_3} \tl n^{-1}
+ n \{ jN^i, e^{\xi \tau_3} \}\tl n^{-1},
\ee
where we used the fact that $N$ and $\tN$ have vanishing Poisson bracket. The first bracket in the right hand side
of \Ref{Xg1} gives the action of the algebra on the Hopf section, which we computed above in \Ref{Lder},
\equ
\{jN^i, n \} \equiv {\cal L}_{\hat \tau^i} n = \big(-\tau^i + N L^i\big)n.
\nequ
Concerning the second term in \Ref{Xg1}, we have from \Ref{PTh},
\be
n \{jN^i, e^{\xi \tau_3} \} \tl n^{-1} = n \{jN^i, \xi \} \tau_3 e^{\xi \tau_3} \tl n^{-1} = 
-L^i \, N g.
\ee
Putting these two together we get the desired result,
\be\label{Xg2}
\{X^i, g\} = (-\tau^i+N L^i)g - L^i \, N g = -\tau^i g.
\ee

Let us also check the action of $\tl X$, the third bracket of \Ref{PT*G1}. We have
\be\label{tlXg1}
\{\tl X^i, g\} = \{ -j\tN^i, n e^{\xi \tau_3} \tl n^{-1}\} = - n \{j\tN^i, \xi \} \tau^3  n^{-1} g
- n e^{\xi \tau_3}\{ j\tN^i, \tl n^{-1} \}.
\ee
Using \Ref{PTh} the first term gives $L^i \, Ng$, whereas the second bracket computes to
\be
\{ j\tN^i, \tl n^{-1} \} =  -\tl n^{-1} \{ j\tN^i, \tl n \} \tl n^{-1} =
-\tl n^{-1} \big({\cal L}_{-\hat \tau^i} \tl n\big) \tl n^{-1} = -\tl n^{-1}\big(\tau^i-\tN L^i\big).
\ee
Using the fact that $g\tN=N g$, we obtain
\be\label{tlXg2}
\{\tl X^i, g\} = L^i \, Ng +g (\tau^i-\tN L^i) = g\tau^i - L^i \left(Ng - g\tN \right) = g\tau^i.
\ee

Finally, one could similarly proceed to check that functions of $g$ commute, although the direct computation via 
Poisson bracket is more intricate.

\item In the course of this work, we realized that the same decomposition \Ref{ThetaTheta} of the $T^*\SU(2)$ symplectic potential was considered long ago by Immirzi \cite{Immirzi}. Here we recover the same result, thus our work on twisted geometries turns out to develop and extend ideas already present back then. In particular, a crucial remark is the following: in \cite{Immirzi} it is argued that the extrinsic curvature should be wholly carried by the extra angle that we denote $\xi_e$, and that this leads to an obvious difficulty, since $\xi_e$ does not have enough degrees of freedom to characterize a discretization of the full extrinsic curvature. The criticism is well-posed, and its solution lies in the fact that part of the extrinsic curvature is carried by the variables $N$ and $\tN$, as already anticipated in the introduction.
We will come back to this important point below in Section \ref{SecGeoInt}.

\end{itemize}

\medskip
Let us summarize where we stand. In this Section, we have introduced the phase space $\bar P$ with Poisson brackets \Ref{PP}, and showed that it is symplectomorphic to $T^*\SU(2)$. The symplectomorphism extends straightforwardly to the whole triangulation, so we conclude that
\equ\label{PcongT}
\bar P_\Ga \equiv \BigTimes_e \bar P_e / \Z^2 \cong \BigTimes_e T^*\SU(2)_e.
\nequ

The symplectomorphism \Ref{PcongT} allows us to give a completely new parametrization
of the kinematical phase space of loop quantum gravity on a fixed graph as the space of twisted geometries. 
This result answers the question raised in the introduction, and shows that there is a natural discrete geometry associated with the space of holonomy-flux variables:
the latter space can be written in terms of areas, normals and an abelian ``connection'' $\xi \in \cs^1$, which can be thought of as the modulus of the extrinsic curvature gauge-fixed as in \Ref{Agf}.
Notice that this discrete geometry is not a Regge geometry, in particular, it is discontinuous because the shapes of the triangles do not match in general.

\section{Abelianization of the gauge-invariant phase space}
The description of the kinematical phase space of loop gravity in terms of the discrete, discontinuous twisted geometries is particularly useful when one works at the gauge-invariant level. To fix ideas, let us consider a closed, 4-valent graph. For this, the familiar relation $2E=4V$ between the total number of edges and that of vertices holds. Then, the Hamiltonian reduction of the phase space gives a dimensionality of $6E- 6V=2E +2V$, which is a pair of conjugate variables for every edge, and a pair of conjugate shape parameters for every vertex.
More  generally, one expects 2 variables for each edge, and $2(n-3)$ variables for each $n$-valent vertex. The question is then how to conveniently extract this set of variables from the initial $(g_e,X_e)$ of the kinematical space.
As we now discuss, twisted geometries give an explicit answer.

The gauge-invariant phase space of loop gravity is obtained imposing the Gauss law constraint at each vertex, and then dividing out the action of the  SU(2) gauge transformation it generates, 
\be\label{Sgi}
\bar S_{\Gamma} \equiv \BigTimes_{e} T^{*} \SU(2)_{e}/\!/ \SU(2)^{V_{\Gamma}}.
\ee
Here $V_{\Gamma}$ is the total number of vertices in the graph. 
In order to impose the Gauss law, there is one important aspect that we need to recall. The kinematical space is given by the assignment of the holonomy-flux variables $(g_e, X_e)$ on a fixed, oriented graph and under reversal of the orientation of an edge, $g_{-e}=g_{e}^{-1}$ and $X_{-e}= - g_{e}^{-1}X_{e}g_{e} \equiv \tl X_e$. Since we are trivializing $T^*\SU(2)$ with right-invariant vector fields, the latter property means that under reversal of the orientation we get the left-invariant one.
Thanks to this fact, the Gauss law can be defined as $C_v := \sum_{e | s(e) =v} X_{e} + \sum_{e | t(e) =v} \tl X_{e} =0$ at each vertex.

Notice the ``non-local'' nature of the quotient in \Ref{Sgi}: each edge subspace is affected by the two vertices it connects.
This is the complicate feature that our new parametrization \Ref{map} simplifies, since we assign to every edge two unit vectors, $ N_{e}$ to the source vertex and $\tilde{N}_{e}$ to the target vertex, and the relation $X_{-e}= - g_{e}^{-1}X_{e}g_{e}$ is automatically taken into acount.  Then the Gauss law can be imposed, and the quotient by $\SU(2)^{V_{\Gamma}}$ can be taken, at each vertex separately.
Furthermore, through the map \Ref{map}, the Gauss law \emph{coincides with the closure constraint} \cite{CF3} (see also \cite{Barrett}),
$C_v := \sum_{e | s(e) =v} j_eN_{e} - \sum_{e | t(e) =v} j_e\tl N_{e} =0$. 
Therefore the procedure amounts exactly to the symplectic reduction already discussed in Section \ref{SecAA}. 

In details, consider the presymplectic kinematical space \Ref{P}, parametrized by the twisted geometries. We can factorize it as a product over edges \emph{and} vertices, analogously to what we did with \Ref{Pev} for the initial area-angle space:
\be
P_{\Gamma}=  \BigTimes_{e} T^{*}S^{1} \BigTimes_{v} \left(\BigTimes_{e\supset v} S^{2}_{j_{e}}\right). 
\ee
We now identify the Gauss law with the closure condition, and take the symplectic quotient locally at each vertex. As already discussed in Section \ref{SecAA}, this amounts to impose the classical closure condition \Ref{closure} and to divide by the $\SU(2)$ rotations it generates. The result on each vertex is the space of shapes of the polyhedron,
$S_{\vj_{v}} \equiv \bigTimes_{e}\cs^2_e /\!/ \SU(2)$, which is a $2(n-3)$-dimensional phase space itself \cite{CF3,FKL,Kapovich}. 
On each edge, although the closure does not affect the $\xi_e$ directly as a constraint, it does as the generator of SU(2) transformations, since
\be\label{xiC1}
\{\xi_e, C_{s(e)}^i \} = L^i(z_e), \qquad \{\xi_e, C_{t(e)}^i \} = -L^i(\tl z_e).
\ee
This double action on each $\xi_e$ has the role of shifting the Hopf sections.
Hence, the reduction requires a \emph{gauge-fixing} of the choice of Hopf sections in the $\xi_e$ variable. Let us assume that the graph is such that this fixing can be done globally without ambiguities.\footnote{An more complicated question is on the other hand to find a \emph{gauge-invariant} reduction of the $\xi_e$ variables. To understand the problem, consider first the simple example of a single edge closed on itself, 
\ \ \begin{picture}(0,0)\put(0,0){\circle*{3}}\put(0,5){\circle{10}} \end{picture} \ \ . In this degenerate case, the unique closure $C=jN-j\tN=0$ implies only the two conditions $\tN = N$, i.e. $\tl z=z$. It is then straighforward to see from \Ref{xiC1} that $\{\xi, C\}$ vanishes on shell, hence $\xi^{\rm g.i.}\equiv \xi$ is a gauge-invariant variable. 
However, this vanishing would be lost had we chosen a non-matching section between $n$ and $\tl n$. In other words, although we are free to choose the sections of the Hopf bundle at the kinematical level, imposing the gauge-invariance removes this freedom. 
Hence, finding a gauge-invariant angle $\xi_e^{\rm g.i.}$ requires also finding a consistent choice of sections throughout the graph.
We leave this issue open for future studies.} Then, denoting $\xi^0_e$ the gauge-fixed variables,
we obtain
 \be\label{SG}
 S_{\Gamma} = \BigTimes_{e} T^{*}S^{1} \BigTimes_{v} S_{\vj_{v}}.
 \ee
This is a factorization of the presymplectic gauge-invariant phase space in terms of a 2-dimensional phase space assigned to each edge, and a $2(n-3)$-dimensional phase spaces assigned to each $n$-valent vertex. 
The procedure is then completed as before, dividing by the kernel of the gauge-reduced symplectic 2-form. This results in a symplectic space $\bar S_\Gamma$ isomorphic by construction to the gauge-invariant phase space \Ref{Sgi}, namely we obtain \Ref{SGamma}.

This decomposition shows that the gauge-invariant space is described by closed twisted geometries, and factorises as a product of phase spaces associated  with edges and vertices. 
This factorization offers a classical analogue to the decomposition $L^{2}(G_{\Gamma}) = \oplus_{j_e} \left(\otimes_v {\cal H}_{\vec{\jmath}_v} \right)$ of the gauge-invariant Hilbert space on a fixed graph of the quantum theory.
In particular, closed twisted geometries realize explicitly the counting described earlier: 2 variables per edge and $2(n-3)$ per vertex. 
The edge variables are still the abelian pairs $(j_e,\xi_e^0)$, whereas the vertex variables are suitable cross ratios parametrized by $n-3$ complex variables $Z_v$ \cite{CF3,FKL}.
For instance in the 4-valent case, the space of shapes of the tetrahedron is two dimensional and can be conveniently parametrized by a complex variable defined as the cross ratio of the stereographic complex coordinates parametrising the 4 points on the spheres meeting at $v$.
Therefore the gauge-invariant phase space can be fully parametrized by abelian complex labels,
\equ\label{abelian}
 S_{\Gamma} = \left\{ Z_e, Z_v\right\} ,
\nequ
where $Z_{e}\equiv j_{e}+i \xi_{e}^0$ is the complex coordinate associated with edges.

This complete factorisation and the related \emph{abelianization} of the loop quantum gravity gauge-invariant phase space is the most remarkable property of the new parametrisation introduced here.
It will play a key role in the quantization of this phase space and in the construction of coherent states \cite{noiQ}.

\section{On the separation between intrinsic and extrinsic geometry}\label{SecGeoInt}

We now want to come back to the interpretation of the variables $(N_e,\tN_e,j_e,\xi_e)$
as extrinsic and intrinsic geometrical data. Thus far, the interpretations of $j_{e}$ as the oriented area of the face dual to $e$, and of $\xi_{e}$ as the norm of the extrinsic curvature integrated along $e$, are clear.  
The question to resolve concerns the interpretation of $N_{e}$ and $\tN_{e}$, and the separation between intrinsic and extrinsic geometry.

The connection between $(g_e, X_e)$ and the intrinsic $E^a_i(x)$ and extrinsic $K_a^i(x)$ geometry is easily done in the continuum limit, where
$g_e \simeq {\mathbbm 1} + A_e$, $X_e \simeq E_e$, and 
\be\label{ThetaLimit}
\f1{\ga}\tr(X_e \rd g_e g_e^{-1}) \simeq \f1\ga E^e_i \rd A_e^i.
\ee
From this $\Omega=-\rd \Theta = \f1\ga \rd A \w \rd E$, and the familiar brackets of loop gravity follow,
\be
\{A^i_e, E^{e'}_j\} = \ga \d^i_j \d^{e'}_e.
\ee
Finally, using 
\equ\label{Aconn2}
A^i_a = \Ga^i_a(E) + {\ga} \, K^i_a,
\nequ 
one recovers the Poisson bracket in ADM variables, $\{K^i_e, E^{e'}_j\} = \d^i_j \d^{e'}_e$.

Here on the other hand we are interested in this separation in terms of some discrete quantities functions of the twisted geometries,  \emph{without taking the continuum limit}. Namely, we want to introduce two quantities ${\bA}_e(N,\tN,j,\xi)$ and ${\bE}_e(N,\tN,j,\xi)$, functions of the twisted geometries to determine, such that splittings like \Ref{ThetaLimit} and \Ref{Aconn2} hold without taking the continuum limit.

To that end, let us first of all define $\bA_e$ simply as the Lie algebra element whose exponential gives the holonomy of the connection \Ref{mapg}, 
$g_{e} \equiv e^{\bA_{e}}$. Then, let us investigate what combination of $N_{e},\tN_{e}$ can be identified as  purely intrinsic geometry, i.e. the combination which (i) Poisson commutes with itself, and (ii) is dual to $\bA_{e}$.
Using the fact that
\be
\rd g_{e}g^{-1}_{e}= J(-{\rm ad}_{ \bA_{e}}) \cdot \rd \bA_{e}  ,
\ee
with $J(x) \equiv (1- e^{-x})/{x}$ and ${\rm ad}_{Y} \cdot X \equiv [Y,X]$, we can write the symplectic potential associated with each edge as follows,
\be
\Theta_{P_e} = \f1{\ga}\tr( X_e \rd g_e g_e^{-1}) = \f1{\ga}
\tr\left( j_e N_e J(-{\rm ad}_{ \bA_e})\cdot \rd \bA_e) \right) = \f1{\ga} \tr\left( \bE_e \rd \bA_e \right),
\ee
where
\be\label{Eintr}
\bE_e \equiv j_e J({\rm ad}_{ \bA_e})\cdot N_e.
\ee
This clearly identifies the intrinsic geometry variables, since it satisfies 
\be
\bE_{-e}=-\bE_{e},\quad \{\bE_{e},\bE_{e'}\}=0, \quad \{\bE^i_{e}, \bA^j_{e'}\}=\ga \d^{ij} \delta_{e,e'}.
\ee
As desired, these variables Poisson commute with each other.

The next step is to identify what is the discrete analogue of the extrinsic curvature, namely we want to introduce further discrete quantities $\bG$ and $\bK$ such that 
\be\label{AGK}
\bA \equiv \bG + \ga\bK.
\ee 
This step more involved, and our argument will be less conclusive.
We have already argued in Section \ref{SecTwi} that one can always choose a gauge such that \Ref{Aconn2} reduces to \Ref{Agf}, and thus $\xi_e$ should be seen as the modulus of $\ga K_e$ in this gauge. From this, we would like to infer that for $\xi=0$ also $K$ and $\bK$ vanish, so that
\be\label{eGamma}
e^{\bG_{e}} = n_{e} \tilde n_{e}^{-1}.
\ee
This would allow us to identify the finite version of the spin connection $\bG$ in terms of the twisted geometries. However, there is a subtlelty: recall in fact that there is a gauge ambiguity $n\to ne^{\alpha(N)\tau_{3}}$ in the choice of the section. So what we mean by the above formula is that we expect that there exists a choice of section $n_{e}, \tilde{n}_{e}$ (this choice might be different for different edges) such that the above interpretation hold.
What this section is can be worked out explicitely \cite{Barrett} in the case the data satisfy the gluing constraints. What section one should choose in the case when these constraints do not hold is not clear, but for the discussion we assume that there is such a choice.

Next, notice that we can write 
\be
g_e =  n_e e^{\xi_e \tau_3} \tilde n_e^{-1} = n_{e} \tilde n_{e}^{-1} e^{\xi_e \tN_e} = e^{\bG_e} e^{\xi_e \tN_e}.
\ee
At this point we use the decomposition of exponentials
\be
e^X e^Y = e^{X + J^{-1}({\rm ad}_X)(Y) + o(Y^2)},
\ee
where
$
J^{-1}(x) = {x}/({1-e^{-x}}),
$
to get
\be
e^{\bG_e} e^{\xi_e \tN_e} = e^{\bG_e + \xi_e J^{-1}({\rm ad}_{\bG_e})\cdot \tN_e + o(\xi_e^2)}.
\ee
We have 
\be
J^{-1}({\rm ad}_\bG)\cdot \tN = J^{-1}({\rm ad}_\bG) e^{{\rm ad}_{\bG}}\cdot N=
J^{-1}(-{\rm ad}_\bG)  \cdot N.
\ee
Thus from \Ref{AGK} we have the identifications
\be \label{Aextr}
 \bG = \ln (n \tilde n^{-1}), 
 \qquad \quad \ga \bK = \xi J^{-1}(-{\rm{ad}}_\Ga) (N) + o(\xi^2).
\ee
Summarizing, under the assumptions (i) of a decomposition \Ref{AGK} at the discrete level, and (ii) that there exists a consistent choice of section realizing \Ref{eGamma}, we are able to identify a combination of twisted geometries describing purely the extrinsic curvature $\bK$, given for small curvature by the second term in \Ref{Aextr}. This shows in particular that the  vectors $N_e$ and $\tN_e$ contain information on both intrinsic \Ref{Eintr}  and extrinsic (\ref{Aextr}) geometry.

Although this discussion is preliminary and needs to be further developed, the important message is clear: 
the vectors $N_e$ and $\tN_e$, \emph{and thus the fluxes}, carry information on the extrinsic curvature. This is also the reason why the fluxes do not commute in loop gravity, because ultimately they not only capture information only about the intrinsic  metric but also about the extrinsic curvature (see also discussion in \cite{AbZapa}).

\subsection{Gluing conditions and Regge phase space}\label{SecRegge}
We have shown that twisted geometries describe a notion of discrete geometry associated to the kinematical phase space of loop gravity, whose intrinsic part is carried by $(j_e, N_e, \tN_e)$, and extrinsic part by $(\xi_e, N_e, \tN_e)$. A similar separation can be made also after the closure condition is imposed over the whole graph, i.e. in the space of closed twisted geometries corresponding to the gauge-invariant phase space of loop gravity. 
This reduced space still describes discontinuous metrics, because of the shape-matching problem discussed in Section \ref{SecAA}. As already pointed out, this problem is caused simply by the fact that $(j_e, N_e, \tN_e)$ carry information about both intrinsic and extrinsic geometry, and thus it cannot be purely interpreted in terms of a three dimensional discrete geometry.

At this point, one might also wonder what happens if the shapes are made to match.
To make the shapes match and the geometries continuous, one needs to add gluing contraints along the lines discussed in Section \ref{SecAA}.  
Since the reduction by the gluing constraints of areas and normals alone corresponds to Regge calculus, the reduction of the closed twisted geometries gives a notion of phase space for Regge calculus, described by the (now continuous) piecewise-flat Regge metrics, and their extrinsic curvature.

As these constraints provide a further restriction than the Gauss law, such a \emph{Regge phase space} is \emph{smaller} than the gauge-invariant loop gravity phase space on a fixed graph $\bar S_\Ga$, a point already made in the literature \cite{BiancaJimmy,Eugenio}. This should not come as a surprise:
each configuration of holonomies and fluxes in $\bar S_\Ga$ corresponds to infinite possible continuous metrics, but in general none of these will be piecewise flat. Such characterization requires additional conditions, which thanks to the twisted geometry parametrization of the gauge-invariant phase space, are manifestly identified precisely by the gluing constraints.

It would be interesting to carry this program of additional reduction further, in particular, to study the relation of our angles $\xi_e$ to the natural variables carrying extrinsic curvature in Regge calculus, namely the four dimensional dihedral angles $\th_e$ (this the angle between the two normals to the tetrahedra sharing the triangle $e$ embedded in the four dimensional spacetime).
We leave this issue open for the moment, but we point out that preliminary conclusions can be drawn from work in the spin foam formalism \cite{CF3,Barrett,Bonzom}. There emerges, at least for the simple triangulation corresponding to the boundary of a 4-simplex, an explicit relation between an angle like $\xi_e$ and the dihedral angle, relation depending on the immersion of SU(2) into the covariant group and thus on the Immirzi parameter. 

Furthermore, the construction of a phase space for Regge calculus has been investigated by Bahr, Dittrich and Ryan \cite{BiancaJimmy,BB} (see also \cite{Waelbroeck}). Their approach differs from ours in being covariant and related to a specific action, thus for instance they use a covariant (Euclidean) SO(4) Poisson structure, and four dimensional normals to tetrahedra. However there are important points in common, such as the basic role of the area-angle variables and the introduction of an angle like $\xi_e$ to encode the connection degrees of freedom, analogue of our \Ref{gNNt}. Because of this, we expect a contact between the Regge phase space they introduce, and the one that can be obtained imposing the gluing conditions on the closed twisted geometries.

\section{Conclusions}

In this paper we presented a new parametrization of the phase space of loop quantum gravity on a fixed graph in terms of quantities describing the intrinsic and extrinsic geometry of a three dimensional triangulation dual to the graph.
The parametrization is based on a symplectomorphism between the holonomy-flux variables, living in the SU(2) cotangent bundle associated with each triangle, and a set of geometric quantities, the \emph{twisted geometries}, that can be interpreted as an assignment to each triangle of its oriented area, two unit normals as seen from the two polyhedra sharing it, and an additional angle related to the extrinsic curvature which parametrises the transformation between the two frames. The Poisson brackets among these variables are given by equations \Ref{PP}, and have an interesting geometric interpretation in terms of a Lie derivative preserving the Hopf section on the sphere.

Initial motivation for these variables comes from the study of coherent intertwiners \cite{LS,CF3}, the construction of the new spin foam models \cite{LS,LS2,FK,Bonzom}, spin foam graviton calculations \cite{grav,gravEPR}, and area-angle Regge calculus \cite{BS,BiancaJimmy,BB}. Here we showed how the variables suggested by these different approaches can be coherently put together to give a new parametrization of the phase space of loop gravity.
The main novelty allowing us to realize this program is the presence of an angle $\xi_e$ per edge, which contains information about the extrinsic curvature and it is essential to reconstruct a parametrization of a discrete version of the classical phase space of general relativity. 

What this parametrization means, is that there is a natural discrete geometry associated with the holonomy-flux variables.
The peculiarity of this twisted geometry is to be \emph{discontinous}, since each triangle has in general a different shape when seen from the two polyhedra sharing it. The geometries can be made continuous imposing suitable gluing constraints, not present in the loop approach, which effectively reduce them to piecewise-flat Regge geometries. This shows the the Regge geometries are a subset of the phase space of loop gravity, a point already made in the literature \cite{Eugenio,BiancaJimmy}. On the other hand, the description in terms of twisted geometries has an intriguing relation to twistors \cite{Twistor}.

A useful property of the parametrization is that the geometric interpretation of holonomies and fluxes given holds for both the kinematical and the gauge-invariant phase spaces of loop gravity. The difference lies in the simple closure conditions being satisfied or not by the variables. When they are satisfied, the variables can be conveniently reduced to a pair of conjugate variables per edge and $2(n-3)$ variables per vertex.
This result provides an abelianization of the gauge-invariant phase space analogue to the decomposition $L^{2}(G_{\Gamma}) = \oplus_{j_e} \left(\otimes_v {\cal H}_{\vec{\jmath}_v} \right)$ of the quantum theory.
This new parametrisation and especially the abelianisation provide a direct and simple route towards quantisation of the loop quantum gravity phase space in terms of coherent states labeled by the twisted geometries.
This will be the subject of a follow up work \cite{noiQ}. 

It should be remarked that similar ideas had been investigated by Immirzi \cite{Immirzi}. In particular, he had already considered the same symplectomorphism, but was worried by how restrictive it looks to describe the extrinsic curvature purely in terms of (his equivalent of) $\xi_e$. As we showed in this paper, the problem does not exist, because the extrinsic geometry is not fully captured by $\xi_e$, but also contained in the normals $N_e$ and $\tN_e$.
Our construction also clarifies the fact that the flux operators $X_{e}$ satisfying the relation $X_{-e}=- g_{e}^{-1}X_{e}g_{e}$ carry both intrinsic and extrinsic geometry.
It is therefore not appropriate to think of them as providing information only about the 
boundary metric, at it is usually done in the LQG literature.
It also explains why this set of operators does not commute, a property that would be puzzling had they been purely intrinsic.

Finally, we also considered the problem of constructing discrete variables, functions of the twisted geometries, corresponding to purely intrinsic and purely extrinsic geometry, and thus representing the original classical algebra.
Such a construction can be viewed as a discrete analogue of the decomposition  $A=\Gamma +\ga K$, which is the cornerstone of the equivalence between $SU(2)$ gauge theory and gravity.
We proposed here an explicit construction which relies on some assumptions, and we believe it would be useful to further investigate this problem, so to fully control the separation at the discrete level of the intrinsic and extrinsic geometry contained in the sphere variable $N_{e}$ and $\tN_e$.

\subsection*{Acknowledgements}
We would like to thank Abhay Ashtekar for encouragement to clarify the global validity of our symplectomorphism, and Carlo Rovelli and Kirill Krasnov for discussions and useful comments on a draft of this paper.


\end{document}